# 3D Guard-Layer: An Integrated Agentic AI Safety System for Edge Artificial Intelligence

Eren Kurshan
*Princeton University*
ekurshan@princeton.edu

Yuan Xie
Hong Kong University
of Science and Technology
yuanxie@ust.hk

Paul Franzon
*North Carolina State University*
paulf@ncsu.edu

*Abstract*—AI systems have found a wide range of real-world applications in recent years. The adoption of edge artificial intelligence, embedding AI directly into edge devices, is rapidly growing. Despite the implementation of guardrails and safety mechanisms, security vulnerabilities and challenges have become increasingly prevalent in this domain, posing a significant barrier to the practical deployment and safety of AI systems. This paper proposes an agentic AI safety architecture that leverages 3D to integrate a dedicated safety layer. It introduces an adaptive AI safety infrastructure capable of dynamically learning and mitigating attacks against the AI system. The system leverages the inherent advantages of co-location with the edge computing hardware to continuously monitor, detect and proactively mitigate threats to the AI system. The integration of local processing and learning capabilities enhances resilience against emerging network-based attacks while simultaneously improving system reliability, modularity, and performance, all with minimal cost and 3D integration overhead.

## I. INTRODUCTION

In recent years, AI systems have been deployed across a wide range of industries and use cases. The use of edge AI grew rapidly across a spectrum of applications from surveillance systems, to autonomous vehicles, medical devices, wireless communication networks, IoT systems and drones [61][62][63]. This surge in adoption has been accompanied with a corresponding rise in the frequency, variety and severity of attacks targeting edge AI systems [56].

Unlike cloud-based centralized AI, edge AI systems face considerable challenges due to the limitations in network bandwidth, computational and learning resources. The growing list of attacks includes adversarial attacks on sensor inputs, such as manipulating images and LiDAR signals to deceive the model, denial-of-service attacks designed to overwhelm the system, sensor jamming that disrupts the system's sensing capabilities through noise and signal interference, data poisoning and others. Due to limitations in system resources and communication bandwidth, such attacks have a more debilitating impact in edge AI systems, potentially leading to catastrophic consequences.

In addition to the growing security challenges, edge AI systems also face an increasingly complex regulatory landscape [65-69] [64]. Scaling the regulatory and audit capabilities to the range and diversity of edge AI systems and tackling the corresponding complexity locally and in real-time is a key obstacle [64].

In this study, we propose an AI safety architecture, 3D Guard-Layer, that employs 3D integration to tackle emerging edge AI safety challenges. This infrastructure provides shadow AI models and local processing capabilities to address resource and communication limitations respectively. It enables continuous safety monitoring along with failover capabilities through shadow AI models.

By processing input and output streams of the primary edge AI system, the 3D Guard-layer allows continuous behavioral and fine-grain hardware monitoring and anomaly detection to identify potential risks or threats. The agentic architecture enables seamless failover capabilities to mitigate ongoing attacks, along with safety and regulatory agents to locally process input streams, preventing downtime and compliance issues during network failures.

In addition to expanding the computing resources in space-constrained embedded applications, 3D architecture offers significant advantages in fine-grained monitoring by fully aligning floorplans and enabling high-degrees of interconnectivity. It requires minimal modifications to the original edge AI chips, enables heterogeneous layer integration, increases transistor density, as well as improving total system power and costs [19][27][29].

This paper is organized as follows: Section II provides background on emerging AI safety and security challenges impacting edge AI systems, Section III describes the attack detection techniques, Section IV discusses the proposed 3D safety architecture and capabilities, Section V describes the agentic AI capabilities, Section VI provides 2.5D and 3D configurations, Section VII describes the cost and overhead analysis for edge AI systems. Section VIII highlights the advantages of the 3D guard-layer architecture, and finally Section VIII provides concluding remarks.

## II. EMERGING EDGE AI VULNERABILITY & ATTACKS

In recent years, the AI attack landscape has evolved rapidly, with an expanding array of increasingly sophisticated threats. This section aims to highlight a selection of recent attack types, illustrating the breadth and diversity of emerging threats with implications on edge AI systems.

Edge AI system architectures typically include multiple layers though the complexity within the layers vary significantly. All layers are subject to a growing number of attacks. *(i) Application Layer:* AI service layer for specific applications such as health, smart vehicles, smart city applications. In more complex edge AI systems, application programming interfaces, middleware, service and device manager layers are included. The safety challenges in this layer may involve API calls to the applications through HTTP protocol. *(ii) Network Layer:* This layer captures network connectivity and vulnerabilities over RFID, WiFi, 4G/5G, Bluetooth through protocols like HTTPS, common IoT protocols like MQTT. DDoS and other network attacks primarily influence this layer through encryption, authentication and other security vulnerabilities. *(iii) Device Layer:* Device layer involves the user and environmental interactions such as sensors, actuators and other data streams. Device or edge layer involves a number of safety vulnerabilities that range from Sybil attacks to jamming and node failure attacks.

### A. Adversarial Attacks on Object Recognition

Attacks on object recognition systems have been among the earliest threats to edge AI systems. These attacks frequently involve introducing subtle perturbations to the data, leading to misclassifications and incorrect decisions. As widely demonstrated in autonomous vehicles, altering the traffic signs through pixel-level perturbations causes the AI system to make incorrect decisions, thereby compromising safety[48].

### B. Spoofing Attacks

Spoofing attacks target edge AI system components such as global navigation satellite system (GNSS), global positioning system (GPS), and Lidars [57][50]. These attacks frequently hijack autonomous vehicles as well as drone systems [49].

### C. Large Language Model Manipulations

The increasing integration of LLMs into edge AI systems have caused a rise in manipulation attacks such as LLM jailbreaking [81], prompt injection [80] in recent years. These attacks direct the edge AI systems to diverge from safe behaviors and by-pass existing guardrails [61].

### D. Multi-Modal Attacks

With the introduction of multi-modal AI systems, developing consistent security measures and guardrails across modalities has emerged as a serious problem. The introduction of a widened attack surface resulted in new and highly effective attack types such as best-of-N (BoN) jailbreaks [47] [52] [72].

### E. Network-based Attacks

Network-based attacks on edge AI include classic attack types like man-in-the-middle attacks [74], denial of service attacks (DDoS) [73], communication interception [31], malware and ransomware attacks [86], compromised software updates [75], unauthorized access through password cracking and others [53][54]. Similarly, Sybil attacks, in which an attacker undermines the network's reputation system through a large number of pseudonymous identities use this to mislead the edge system and direct it to behave abnormally have significant implications for edge AI systems [58].

### F. Data Poisoning Attacks

Though it has characteristics similar to the network-based attacks, data poisoning is emerging as a growing category of its own [77]. Through data poisoning attacks, the edge system is fed poisoned data such as false sensor data to lead to unsafe decisions and manipulation by the adversaries [55].

### G. Hardware Attacks on Edge AI

Hardware attacks, such as electromagnetic fault injection attacks, voltage glitching, hardware Trojans, and side-channel attacks, are significant threats to edge AI systems. While EMFI and voltage glitching can induce faults by manipulating electromagnetic fields and power supply to produce incorrect results or manipulate the inference process in deep neural networks respectively [80], hardware trojans, inserted during manufacturing, can alter system behavior when triggered [81]. Similarly, side-channel attacks exploit physical emissions, such as power consumption or electromagnetic radiation, to extract sensitive information from AI models [82].

## III. ATTACK DETECTION TECHNIQUES

Though the attack types vary significantly attack detection techniques frequently utilize common capabilities such as cross-verification, ensembling and shadow processing, anomaly detection

Object recognition attacks are detected through a number of techniques. Ensemble methods, utilizing multiple classifiers to cross-verify results, facilitate the detection of attacks by identifying inconsistencies across models. Input validation techniques, including preprocessing filters and robustness testing, enable the identification of perturbations in images. Anomaly detection, through statistical modeling of image patterns and tracking of input provenance, further supports the recognition of tampered inputs.

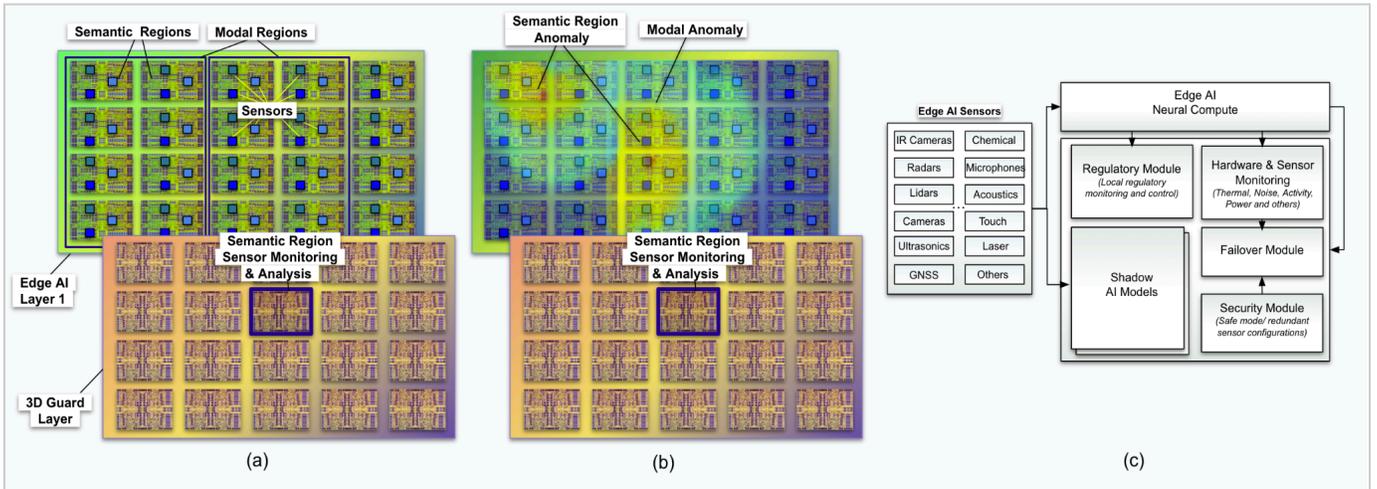

Fig.1 (a) Sensor infrastructure on edge AI layer (top) with semantic, feature and modal analysis, sensor analysis and anomaly detection region in 3D Guard-layer (bottom) (b) Semantic, feature and modal anomaly in edge AI layer (c ) High-level architecture of the 3D Guard-Layer with shadow AI models, dedicated regulatory, security and failover modules as well as hardware and sensor monitoring. The guard layer directly monitors all layers

Similarly, spoofing attacks can be detected through anomaly detection models to identify deviations from typical input patterns, flagging suspicious data. Sensor cross-validation across multiple sensors, such as cameras and accelerometers, help identify inconsistencies that may indicate spoofing. Additionally, behavioral analysis of interaction patterns can detect unusual user or device behavior, further enhancing the ability to identify malicious activity. These methods work together to improve the robustness of edge AI systems against spoofing attacks.

LLM attack detection involves several techniques, including input anomaly detection, where unusual text patterns or statistical deviations from normal inputs are flagged. Behavioral monitoring identifies inconsistencies or harmful outputs, such as contradictions or biased responses, indicative of adversarial manipulation.

Network attacks on edge AI systems are frequently detected using a combination of anomaly detection, intrusion detection systems, and behavioral anomaly detection. Anomaly detection compared to baselines of normal network traffic, allows the detection of deviations indicative of potential attacks. In addition, some of the network based attacks also yield hardware behavior anomalies, such as overactivity in communication modules, abnormal processing patterns in malware, ransomware and DDoS attacks etc.

## IV. 3D GUARD-LAYER FOR EDGE CO-PROCESSING

This section describes the 3D Guard-layer architecture aiming to improve the overall safety of edge intelligence systems. The 3D Guard-layer is a vertically integrated layer within the primary edge AI system such that it provides: (i) real-time local monitoring, (ii) shadow processing, (iii) failover and (iv) regulatory processing capabilities.

The 3D Guard-layer architecture aims to provide the basic hardware infrastructure support to detect and remediate a wide range of attacks enabling a wide range of safety monitoring, attack detection and remediation techniques. 3D Guard-layer variants can be further customized for specific application families and integrated in a modular fashion.

*A. Hardware Infrastructure for Real-Time Monitoring and Anomaly Detection*

1) System-level Behavioral Monitoring: 3D Guard-layer architecture uses a shadow AI layer to continuously monitor the behavior of the primary AI chip(s) for safety threats and behavioral anomalies. Co-location of the guard-layer provides direct access to the activity on primary computing layers for anomalous behavioral pattern detection. As an example, in an autonomous vehicle use case driving patterns may be tracked to detect anomalous events by the shadow layer.

2) Hardware Monitoring: The system leverages the sensor monitoring infrastructure on the main AI chip for hardware monitoring. This may further be enhanced by additional sensor support on the primary AI system such as block-level activity monitors, power, noise and thermal sensors. As shown in Fig.1(a) the semantic tiles incorporate hardware sensors in the primary edge AI layer(s), which are tightly interconnected through TSV and uC4 to the adjacent 3D Guard-layer with sensor monitoring and anomaly detection regions. Common attack types such as DDoS, flooding and malware attacks can be detected through the anomalous thermal sensor readings and activity monitor patterns.

3) Semantic Monitoring: The popularity of architectures with semantic specialization such as YOLO [26], Deeplab [11], faster R-CNN [96] and others provides the opportunity to monitor semantic block, region level and modal monitoring to detect anomalies that involve activations of incorrect modalities or semantic regions during attacks. As shown in Fig.1(b) anomalies in semantic regions or modalities are picked up by the sensor networks in edge AI layers and the semantic monitoring and analysis regions in 3D Guard-layer.

Recent attack types can be detected through anomalous computational demands or excessive activity in specific regions. For example, multimodal jailbreaking induces sudden spikes in computational load in specific modalities, while poisoning attacks result in abnormal computational demands across the system [72]. Indirect prompt injections embedded in image or audio inputs often lead to unexpected and disproportionate activity within the corresponding modal regions [78]. Poisoning, malware, and takeover attacks induce abnormal behaviors in unusual regions. While anomalous patterns exhibit in a range of hardware configurations, semantic monitoring provides more insights in neuromorphic hardware. Similarly, the infrastructure enables feature analysis by examining activation patterns and conducting feature space comparisons to identify discrepancies in the recognition process.

*B. Hardware System Support for Shadow Processing to Mitigate Attacks and Failover*

In the event that an on-going attack or a security compromise is detected, the 3D guard-layer provides infrastructure support for remediation techniques.

1) Shadow AI Models: Shadow AI models are integrated in the 3D security layer serving functions such as:

*(i) Shadow Co-Processing for Safety*: In some configurations shadow models use alternative redundant sensors for co-processing to detect anomalous behaviour. As shown in Fig.2 the 3D Guard-layer architecture includes monitoring, failover, security and regulatory modules as well as shadow AI models that input sensory data from sensor devices associated with the edge AI system. For higher-criticality events, and hardware attacks the intervention is shadow chip taking over the decision making processes over the primary chip. As edge AI systems are equipped with redundant sensors, the system reconfigures the use of active sensors to mitigate the on-going attack.

*(ii) Confidence Scoring:* Stochastic nature of large language models present significant safety challenges, which the 3D Guard-Layer architecture addresses by utilizing shadow and ensemble models to detect inaccuracies through entropy-based divergence quantification. Specifically, semantic entropy can be employed to estimate the entropy of meaning distributions in free-form responses, enabling the identification of incorrect or low confidence outputs [59].

*(iii) Learning Attack Patterns:* Shadow AI models provide continuous learning of attack patterns over time as well as coordinating with backend cloud-based AI systems to track novel attack types, security vulnerabilities and remediation techniques to be implemented directly in the learning hardware. Privacy preserving federated learning solutions are also enabled through the distributed learning infrastructure.

2) Failover Capabilities: The 3D Guard-layer architecture includes failover capabilities through a dedicated failover agent. In the event of a model compromise, shadow models are used to detect model compromise through anomalous pattern analysis. The corresponding agent changes the system configuration to replace the primary edge AI model(s) with shadow models to thwart the ongoing attack. This capability supports full offline processing capabilities for DDoS and other network based attacks.

*C. Regulatory Processing AI Agents*

With the introduction of new AI regulations around the globe, the need to perform real-time regulatory monitoring has become a key capability for AI systems. Due to their mobility, edge AI systems may be used in and move through a wide range of jurisdictions with different regulatory requirements. The 3D Guard-layer provides real-time regulatory monitoring, audit and intervention capabilities for both AI and application-specific regulations through shadow processing and continuous monitoring. While the primary model(s) process the data and make decisions, shadow regulatory agent(s) continuously monitor and remediate potential compliance issues.

*D. 3D Configurations*

The proposed system can be implemented through (i) silicon interposers and (ii) 3D configurations to capture the varying degrees of complexity of edge AI systems. Current edge AI systems span the full range from single chip systems to highly complex multi-chip modules. Moreover, 3D ICs are commonly used in edge AI scenarios. This provides an opportunity to integrate 3D Guard-layer with the existing 3D stacks. The 3D floor plan alignment and co-location provides advantages in seamless failover mode activation, shadow security processing, and fine-grain sensory analysis. Through-silicon-via or micro-C4 based access to the main processor layer provides monitoring capabilities at a fine grain. The 3D Guard-layer can be integrated through face-to-back (guard-on-top/edge-on-top) and face-to-face configuration (as shown in Fig.2). 2.5D interposers provide support for existing multi-chip modules (Fig.3).

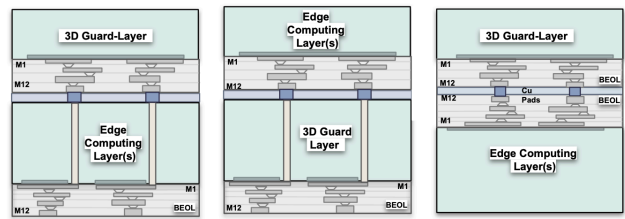

Fig.2 (i) Face-to-Back 3D Guard-layer configuration with edge intelligence layer codesign (ii) Face-to-Back configuration with limited changes to edge intelligence layers (iii) Face-to-Face Configuration

As shown in Fig. 3 the edge AI chips can be integrated with the guard-layer interposer for sensor-based monitoring, passive and active redundant models, and failover hardware.

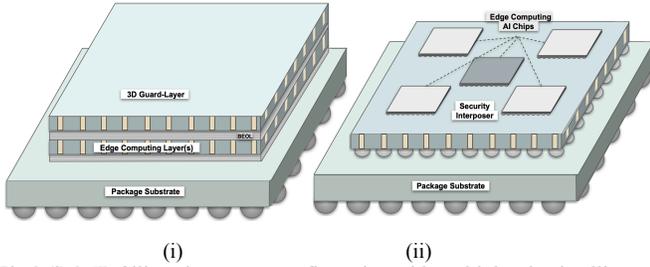

Fig.3 (i) 2.5D Silicon interposer configuration with multiple edge intelligence chips and a dedicated Guardlayer interposer connected to all AI chips (ii) 3D Guard layer configuration with one or more edge intelligence layers and stacked 3D guardlayer.

## VI. AGENTIC AI IMPLEMENTATION

This section presents the agentic implementation of a 3D GuardLayer configuration. The framework incorporates 5 agents: (i) Behavioral monitoring agent, (ii) Hardware monitoring agent, (iii) Shadow Processing Agent, (iv) Failover System Agent, (v) Regulatory Compliance Agent. The agentic AI framework details are described below.

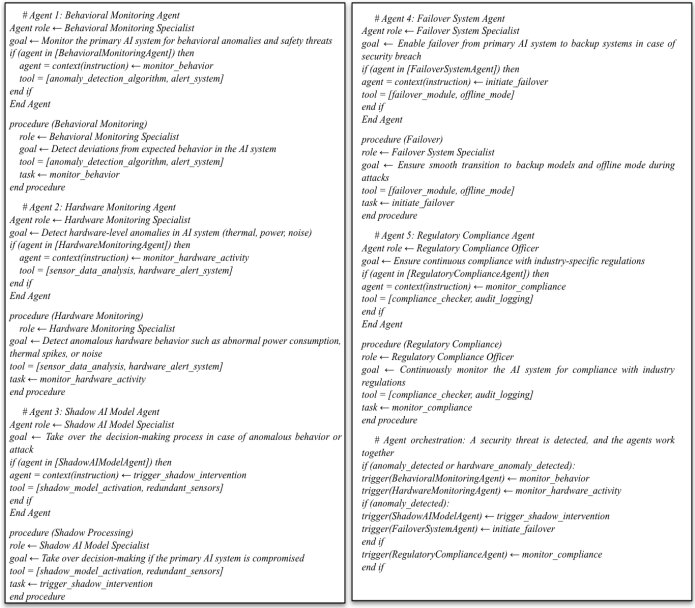

Fig.5 Agentic AI implementation of the 3D Guard-layer system.

## VII. 3D GUARD-LAYER SYSTEM-LEVEL BENEFITS

The proposed 3D Guard-layer architecture provides a number of key system-level benefits in edge AI systems. This section discusses a selection of 3D Guard-layer advantages.

### A. Real-Time Response to Attacks

Current edge AI systems primarily rely on cloud-based backend systems for security and safety matters as they possess limited safety capabilities beyond encryption due to the local resource limitations. The resulting reliance on the cloud systems and communication networks introduces associated delays, thereby limiting the AI system's response time to emerging attacks. It increases the safety risks and catastrophic failures due to these extended response times. The 3D Guard-layer architecture provides real-time local detection and safety capabilities as well as protection from network failure and network related attacks.

### B. Reduced Network Attack & Failure Risk

3D Guard-layer reduces communication delays and demands by locally processing safety related tasks. Given that a large percentage of edge attacks exploit communication channels, this improves safety and provides resilience against network attacks Moreover, the system exhibits enhanced resilience to network failures and outages, improving the operations and safety during such events.

### C. Improved Privacy & Data Protection

Network-based attacks also pose significant threats to data protection and privacy. The local data processing reduces the need to transmit sensitive data to centralized servers. This results in improved data privacy and protection which is critically important in high-security use cases such as medical applications.

### E. Modularity & Customization of Local Safety Capabilities

Edge AI systems span a broad spectrum of applications making them vulnerable to a variety of attack types. However, customizing edge AI systems for specific attack types for an application is costly, which results in reliance on centralized cloud infrastructures. The proposed system offers a cost-effective solution for integrating custom AI safety features into edge devices, tailored to diverse operational and environmental conditions. By utilizing baseline 3D infrastructure or silicon interposer support, a range of safety capabilities can be seamlessly incorporated into individual edge AI systems.

### F. Deep Inspection of Attacks

3D stacking enables fine-grain monitoring of the underlying computation in the hardware at a block or macro-level through a high-bandwidth interconnect to power, thermal, and activity sensors. This provides deeper insights on the source and semantic details of the attacks, modality of the attacks, risky objects for recognition attacks and other characteristics.

### G. System Reliability and Performance Improvement

The 3D Guard-layer serves as an embedded redundancy for learning hardware attacks and failures by enabling the redirection of the computation from the AI layer to shadow guard layer or layers. This improves system-level reliability, robustness - key to applications like cybersecurity, medical systems, military systems and other mission critical system infrastructure. Furthermore, 3D Guard-layer resources can be used to boost performance in adversarial environments, highly unpredictable scenarios with changing system demands.

*H. Resilience Against Hardware Attacks*

Hardware-based attacks, such as electromagnetic fault injection, hardware Trojans, voltage glitching, and side-channel analysis, have proven effective in compromising edge AI systems. Guard-layer hardware provides the ability to shield the critical device layers from electromagnetic radiation commonly used in trojan triggering, EMFI, side channel and other attacks. The 3D vertical power grid enables protection from power glitching attacks.

*I. Post-Quantum Hardware Demand and Side-Channel*

Increased computational demands of post-quantum cryptography techniques such as recently standardized Dilithium and Kyber, increases the demand for hardware support on the device. Furthermore, these algorithms require additional protections due to their underlying vulnerabilities to side-channel attacks [10] [38] [54]. The side channel vulnerabilities of PQC standards like Kyber and Dilithium due to their lattice characteristics include power analysis, timing analysis and memory access patterns, the 3D guard layer becomes an essential component for hardware support for advanced cryptography techniques.

VII. COST & OVERHEAD ANALYSIS

This section presents a cost and overhead analysis for the 3D Guard-layer architecture and its possible configurations. As discussed earlier, the 3D Guard-layer can be implemented in a range of baseline edge AI systems with single chip, multi-chip, 2.5D and 3D packages. The system itself provides flexible implementation through 2.5D silicon interposer, face-to-face and face-to-back 3D stacking. Table 1 shows a list of commonly used commercial edge AI systems with a range of CPU, GPU, TPU, AI accelerator and neuromorphic architectures.

In terms of 3D vertical interconnect overhead, 75% of the configurations, namely (i) 2.5D, (ii) 3D face-to-face and (iii) 3D face-to-back edge-on-top configurations do not incur any TSV overhead. While these configurations are preferred, TSV overhead for the face-to-back guard-on-top configuration can be estimated as shown in Table 1. Using state-of-the-art commercial 3D technology assumptions such as 3μm size and 10μm pitch, Table 1 provides an interconnectivity sweep from 10K to 100K.

| System | Package | Architecture | Area (mm²) | 10K TSV % | 20K TSV % | 50K TSV % | 100K TSV% | Tech Node |
|---|---|---|---|---|---|---|---|---|
| System 1 [93] | MCM | GPU + CPU | 5133 | 0.00% | 0.00% | 0.01% | 0.01% | 8nm |
| System 2 [94] | MCM | GPU + CPU | 3107 | 0.00% | 0.00% | 0.01% | 0.02% | 8nm |
| System 4 [96] | SCM | Neuromorphic | 400 | 0.02% | 0.04% | 0.09% | 0.18% | 7nm/ 16nm |
| System 5 [97] | MCM (3D*) | FPGA + CPU + AI Engines | 2025 | 0.00% | 0.01% | 0.02% | 0.04% | 7nm |
| System 6 [98] | SCM | Hybrid (CPU, GPU, DLAs) | 289 | 0.02% | 0.05% | 0.12% | 0.24% | 16nm |
| System 7 [99] | SCM | CPU + GPU + NPU | 196 | 0.04% | 0.07% | 0.18% | 0.36% | 14nm |
| System 8 [100] | SCM | Neuromorphic | 800 | 0.01% | 0.02% | 0.04% | 0.09% | 12nm |
| System 9 [101] | MCM/ SCM | ASIC (Tensor Processing Unit) | 150 | 0.05% | 0.09% | 0.24% | 0.47% | 12nm |

Table.1 Through-silicon-via overhead estimations for the only configuration with overhead (F2B 3D configuration) for commonly used edge AI hardware systems in the market

3D manufacturing cost overheads depend on the baseline 3D edge AI system hardware manufacturing costs. The original system costs for the most common edge AI hardware systems exhibit a 20x variation. Similarly, these baseline system packages span multi-chip modules (MCM) and single-chip modules (SCM). As discussed earlier, using the 2.5D guard-layers for MCMs and 3D for SCMs, 3D guard-layer cost overheads range from 3.75% for high-end to mid-range edge computing systems to close to 60% in the lowest cost options due to the low price point of the baseline system. With the rapidly growing complexity of edge AI systems, this manufacturing cost overhead is likely to decline.

VIII. CONCLUSIONS

AI is rapidly adopted in edge computing systems, which caused a similar growth in the number and diversity of attacks. Edge AI hardware and communication resources already overutilized by rapidly growing AI demands face compounding challenges from safety risks and emerging attacks.

This paper proposes a novel agentic AI architecture, 3D Guard-layer, that utilizes 3D integration and agentic AI capabilities to tackle the critical issues in edge AI safety. It introduces key infrastructure capabilities, fine-grain real-time monitoring shadow AI models, offline/failover mode detect and mitigate a wide range of attacks. The infrastructure also provides benefits such as improved system reliability, modularity, speed and performance improvements with minimal cost and 3D overheads.